\documentclass[12pt]{article}
\usepackage[letterpaper,body={8in,11in},margin=1in]{geometry}
\usepackage{ucs}
\usepackage[utf8x]{inputenc}
\usepackage[T1]{fontenc}
\usepackage{ae,aecompl}
\usepackage{lmodern}
\usepackage{microtype}
\usepackage{graphicx}
\usepackage{color}
\usepackage[american]{babel}
\usepackage{xspace}
\usepackage{setspace}
\usepackage[colorlinks=true,urlcolor=blue]{hyperref}
\usepackage{hyperref}
\hypersetup{
  pdftitle    = {Spontaneous Symmetry Breaking},
  pdfauthor   = {Gerald S. Guralnik <gerry@het.brown.edu>},
  pdfsubject  = {High Energy Physics - Theory},
  pdfcreator  = {LaTeX2e with hyperref package},
  pdfproducer = {pdflatex},
  pdfkeywords = {Quantum Field Theory, Schwinger-Dyson Equations, Lattice,
    Symmetry Breaking, Non-Perturbative QFT},
  pdfview     = {FitH}
}
\usepackage{fancyhdr}
\fancypagestyle{plain}{%
\fancyhf{} 
\fancyfoot[C]{\bfseries\thepage} 

}

\newcommand{\helv}{%
\fontfamily{phv}\fontseries{b}\fontsize{9}{11}\selectfont}
\usepackage{amsmath,amsfonts,amstext,amssymb,amsbsy,amsopn,amsthm,eucal}
\DeclareMathAlphabet{\mathpzc}{T1}{pzc}{m}{it} 
\usepackage{mathrsfs}
\usepackage{wasysym}
\usepackage{empheq}
\usepackage{dsfont}
\usepackage{pifont}
\makeatletter
\def\equalsfill{$\m@th\mathord=\mkern-7mu
  \cleaders\hbox{$\!\mathord=\!$}\hfill
  \mkern-7mu\mathord=$}
\makeatother
\def\hksqrt{\mathpalette\DHLhksqrt}
\def\DHLhksqrt#1#2{\setbox0=\hbox{$#1\sqrt{#2\,}$}\dimen0=\ht0
  \advance\dimen0-0.2\ht0
  \setbox2=\hbox{\vrule height\ht0 depth -\dimen0}%
{\box0\lower0.4pt\box2}}

\newcommand{\pa}{\partial}
\newcommand{\comm}[2]{\left[\,#1,#2\,\right]}
\newcommand{\ev}[1]{\left\langle #1\right\rangle}

\newcommand{\ket}[1]{\left| #1\right\rangle}

\newcommand{\ipop}[3]{\ensuremath{\langle#1 | #2 | #3\rangle}\xspace}

\newcommand{\bo}{\raise+0.0mm\hbox{$\Box$}}

\begin{document}
%
\begin{titlepage}
  \begin{flushright}
    {BROWN-HET-1555}
 \end{flushright}
  \bigskip

  \begin{center}
    {\LARGE \textbf{\textsf{The History of the Guralnik, Hagen and Kibble development of the Theory of
          Spontaneous Symmetry Breaking and Gauge Particles}}} \\
    \bigskip

    {\Large \textsf{Gerald S.
        Guralnik\footnote{\href{mailto:gerry@het.brown.edu}{\texttt{gerry@het.brown.edu}}}}} \\
    \bigskip

    \textsf{Physics
      Department\footnote{\href{http://www.het.brown.edu/}{\texttt{http://www.het.brown.edu/}}}.}\\
    \textsf{Brown University, Providence --- RI. 02912}
    \date{\today}
  \end{center}
  \bigskip

\begin{abstract} \noindent I discuss historical material about the beginning
  of the ideas of spontaneous symmetry breaking and
  particularly the role of the Guralnik, Hagen Kibble paper in this development.
 I do so adding a touch of some more modern ideas about the
  extended solution-space of quantum field theory resulting from the intrinsic
  nonlinearity of non-trivial interactions.
\end{abstract}

  \tableofcontents
\end{titlepage}
%
\section{Introduction} \label{sec:0}

This paper is an extended version of a colloquium that I gave at Washington
University in St. Louis during the Fall semester of 2001. But for minor
changes, it corresponds to my recent article in IJMPA \cite{gg;2009}. It
contains a good deal of historical material about the beginning of the
development of the ideas of spontaneous symmetry breaking as well as a touch
of some more modern ideas about the extended solution-space (also called
vacuum manifold or moduli space) of quantum field theory due to the intrinsic
nonlinearity of non-trivial interactions \cite{ggg;1996,ggzg;2007}. While I
have included considerable technical content, a reader interested only in the
historical material should be able to follow the relevant content by just
ignoring the equations. The paper is written from a very personal point of
view. It puts particular focus on the work that I did with Richard Hagen and
Tom Kibble (GHK) \cite{ghk;1964}. Our group and two others, Englert-Brout (EB)
\cite{eb;1964} and Higgs (H) \cite{phpl;1964,ph;1964} worked on what is now
known as the ``Higgs'' phenomenon. I discuss how I understand symmetry
breaking (a viewpoint which has changed little in basics over the years) and
cover the evolution of our 1964 (GHK) work, which was done in its entirety
without any knowledge of others working on the same problem of symmetry
breaking and gauge systems. Later in this document, I will make a few brief
statements accentuating the differences between our work and that of EB and H.

An intense collaborative effort between Hagen, Kibble and me enabled
us to produce a paper that to this day gives us great
satisfaction. The physics I am about to describe was very exciting.
Although we knew the work was basic, we had no appreciation as to how
important the concepts involved would become.  Despite the current
significance of our work, to me the most important thing that has come
from it is my enduring long friendship with Dick and Tom. We came
together to do this work because of fortunate overlaps of interests
and geography. Hagen had been my friend and physics collaborator since
our undergraduate days at MIT.  I went to Harvard for graduate school,
while Hagen stayed at MIT. At that time, in practice, there was little
difference in the training a particle theorist received at either of
these universities. We could easily take courses at either place and
we were constantly doing the short commute between the schools to
``cherry pick'' course offerings. Many of us from Harvard attended a
particle phenomenology course taught by Bernard Feld at MIT. Also, in
the mix, was the painful but very important year-long mathematical
methods course taught at MIT by Feshbach using his book with
Morse. Almost all of us, Harvard or MIT, attended Schwinger's field
theory courses at Harvard. In addition, I took the field theory course
at MIT taught by Ken Johnson, who was Hagen's thesis advisor. This
course was beautifully done and more calculational in nature than
Schwinger's courses. I also sat in on Weisskopf's nuclear physics
course at MIT, which was really fun and, in fact, largely taught by
Arthur Kerman. Of course, I took most of the ``standard'' graduate
courses at Harvard.  Roy Glauber and Wally Gilbert taught courses that
were so well reasoned that I find my notes from their classes still
useful.

Hagen and I wrote our first paper \cite{gh1;1963}
together when we were graduate students, and we continued to talk about life and physics after he went
to Rochester as a postdoctoral fellow, while I concluded my thesis on
symmetry breaking and spin-one fields with primary emphasis on Lorentz
symmetry breaking in four fermion vector-vector couplings \cite{ggff1;194,ggff2;1964}. I was
working under the direction of Walter Gilbert who, by this time, had
largely switched over to biology (Nobel Prize in Chemistry, 1980). Early in 1964, I passed the thesis
exam and with Susan, my new wife, and  a NSF postdoctoral fellowship went, in February, to
Imperial College (IC) in London, where I immediately met Tom and soon began the
discussions that eventually led to our three-way collaboration and
the GHK paper.

While I recount the history of our work, I will do so embedding a fair
amount of physics reflecting how we understood symmetry breaking. Even
after all these years, I feel our understanding is still appropriate
and indeed recently it helped form the basis of what I believe is a
fundamental approach to understanding the full range of solutions
(solution-space  --- this used to be called ``vacuum manifold''; however,
nowadays it is generally referred to as ``moduli space'') of quantum
field theories \cite{ggg;1996,ggzg;2007}. Some constructs from these
modern papers are used in the following discussion to help clarify my points.

\section{What was front-line theoretical particle physics like in the
    early 1960s?} \label{sec:1}
  To set our work in perspective, it is helpful to review what high energy
  physics was like when this work was being done. Many ``modern'' tools such as the
  Feynman path integral and the now nearly forgotten Schwinger Action
  Principle were already available. However, the usual starting point for any
  theoretical discussion was through coupling constant perturbation theory.

  The power of group theory beyond $SU(2)$ was just
  beginning to be appreciated, and was realized through flavor $SU(3)$ as introduced by
  Gell-Mann and Ne'eman. The $\Omega^{-}$, needed to fill in the baryon decuplet (10
  particles) was found in 1963. The Gell-Mann--Zweig quark (ace) ideas had just
  been formulated, but were far from being completely accepted. There was no
  experimental evidence for quarks, and the ideas about color that allowed three quarks to make a
  nucleon only began to take form in 1964 \cite{greenb;1964}. Calculation
  methods were manual and limited. For the most part, coupling constant
  perturbation was the only tool available to try to get valid quantitative
  answers from quantum field theory. Because of the doubts
  that field theory could ever move beyond perturbation theory, $S$-Matrix theory
  was king, at least far west of the Mississippi River. Current algebra, a mixture of
  symmetry and some dynamics was beginning to take shape with work by Nambu and Lurie
   \cite{nl;1961}, but was still in the wings.  Indeed, starting in the 1960s, several of the major
  contributions to what was to become a dramatic reinvigoration of quantum field
  theory came from the works of Nambu and collaborators.

  The Nambu--Jona-Lasinio model \cite{Nambu:1961tp,Nambu;1961tp2}
  described by the interaction:

  \begin{equation*}
    g\, \left[ (\bar{\psi}\psi)^2 - (\bar{\psi}\gamma_{5}\psi)^2 \right] \; 
  \end{equation*}
  played a key role in the development and the understanding of spontaneous
  symmetry breaking in quantum field theory.  There is no bare mass term in
  this interaction and consequently the action has a conserved chiral
  current. This interaction, in itself, is disturbing (as were the
  four-fermion interactions used to describe weak processes) because
  perturbation theory in $g$ produces a series of increasingly primitively
  divergent terms, making this expansion unrenormalizable. Nambu and
  Jona-Lasinio studied this model by imposing a constraint requiring that
  $\langle\bar{\psi}\psi\rangle \neq 0$, which induces a non-vanishing fermion
  mass, and thus seems inconsistent with (or ``breaks'') chiral
  symmetry. Based on this assumption, they formulated a new leading order
  approximation (not a coupling constant perturbation theory). This
  approximation is, in fact, consistent with the conservation of the chiral
  current, thereby assuring that, despite the induced mass, the approximation
  remains consistent with the basic requirements of the field equations for
  this action.

  In addition, their results showed that there was a zero mass composite
  particle excited by $(\bar{\psi}\gamma^5\psi)$. This particle is now called
  a Goldstone boson or occasionally a Nambu-Goldstone boson. The name was
  acquired after developments in other papers \cite{jg;1961,gsw;1962}. Stated
  somewhat more generally, Ref \cite{gsw;1962} proves that if the commutator
  of a conserved charge (resulting from a continuous current) with a local
  field operator has non-vanishing vacuum expectation value, then the local
  field operator must have a massless particle in its spectrum. This result is
  exact and not confined to a leading order approximation. This is the
  Goldstone theorem and it is always true, provided its basic assumptions,
  outlined carefully in what follows, are satisfied.

  The Nambu--Jona-Lasinio approximation can be extended in a consistent way to a well
  defined expansion (all orders, but not in coupling $g$), much as is done in
  Refs \cite{ggff1;194} and \cite{ggff2;1964}. The resulting expansion can
  be renormalized, and the resulting theory looks like a sum of conventional
  theories involving pseudoscalar and scalar couplings to fermions. However,
  the underlying fermionic mass in this  ``phenomenological'' theory is
  generated by symmetry breaking information carried in the vacuum state, and the
  pseudoscalar boson is constrained to be massless.

 \section{A simplified introduction to the solutions and phases of quantum field
   theory } \label{sec:2}

 Before presenting a solution to symmetry breaking quantized scalar
 electrodynamics, the core component of the unified electroweak theory, it is
 worth a short review of properties that are found in solutions to any quantum
 field theory. Considerable insight into the nature of spontaneous symmetry
 breaking in QFT is available through the examination of the solutions of
 comparatively simple differential equations. Some of the material in this
 section, is relatively new and was not available at the time of publication
 of the GHK paper \cite{ggg;1996,ggzg;2007}. In fact, the development of
 these clarifying ideas was inspired by the GHK paper.

A QFT is described by assuming a form for an action. From the action, all
solutions and their Greens functions and hence all properties of the QFT can
be calculated. One of two equivalent methods, the Schwinger action principle
or the Feynman path integral, is commonly used to formulate the equations used
for explicit calculations.  Of course, a major part of a physics problem is to
guess a form for the action which is consistent with known physics and
predicts correct new physics. The basic points that we want to emphasize here can be
demonstrated by looking at the action for a single scalar field $\phi$
interacting quartically with itself and linearly with an external source
$J(x)$ in $4$ space-time dimensions. The (Euclidean) action is given by:

\begin{equation*}
  \int d^4x \bigg[ \phi(x)\,\frac{(-\bo + m^2)}{2} \, \phi(x) +
  g\,\frac{\phi^4(x)}{4} -J(x)\phi(x)\bigg] \; . 
\end{equation*}
The (Euclidean) Schwinger Action principle:

\begin{equation*}
  \delta \left\langle t_1 \right|\left.t_2 \right\rangle = 
  \left\langle t_1 \right|\delta S\left|t_2 \right\rangle \; ,
\end{equation*}
results in the equation:

\begin{equation*}
  (-\bo + m^2)\, \phi(x) + g\, \phi^3(x) = J(x) \; 
\end{equation*}
Defining $\mathcal{Z}$ as the matrix element of a state of lowest energy in the presence
of the source at very large positive time measured against the ``same state''
at very large negative time and again using the Schwinger action principle
leads to:

\begin{equation*}
  \Biggl[(-\bo_x + m^2)\,\frac{\delta}{\delta J(x)}  +
    g\,\Bigl(\frac{\delta}{\delta J(x)}\Bigr)^3 - J(x)\Biggr] \mathcal{Z}[J] = 0 \; .
\end{equation*}
It is important to observe that this method removes all operators and produces
number valued equations. Any Green's function can
be calculated by taking functional derivatives with respect to the source
$J(y)$ of the above equation.

A good way to make sense of this equation is examine it on a space time
lattice with $N$ space time points. This approach can be regarded as the
original definition of a quantum field theory which is realized only in the
limit of vanishing lattice spacing.  On a hyper-cubic lattice:

\begin{equation}
  \bo\ {\phi}_n = \sum_k \left({\phi}_{n+\hat{e}_k} + {\phi}_{n-\hat{e}_k}-2{\phi}_{n}\right),
\end{equation}
where $\hat{e}_\mu$ is a unit vector pointing along the $k$-direction and, for
convenience, the lattice spacing is set to 1. Since functional
derivatives become ordinary derivatives at a lattice point the equation for $\mathcal{Z}$ on the
lattice is:

\begin{equation}
 \Biggl[- \sum_k \left({\phi}_{n+\hat{e}_k} +
   {\phi}_{n-\hat{e}_k}-2{\phi}_{n}\right) + m^2\, \frac{d}{dJ_{n}} + g\, \Bigl( \frac{d}{d J_{n}} \Bigr)^3 - J_{n} \Biggr]\,
    \mathcal{Z}[J_{1},J_{2},\dotsc,J_{p},\dotsc] = 0 \; .
\end{equation}

The space-time derivatives have served to make this an equation involving
three lattice points with the functional derivatives becoming normal
derivatives acting on the variable at the central lattice point. This lattice
equation makes it clear that $\mathcal{Z}$ is described by $N$ linear third order coupled
differential equations in the source $J(x)$. Generally this can be expected to
result in $3\, N$ independent solutions. Many interesting things occur in the limit
of taking an infinite number of lattice points while also moving the lattice
spacing to zero to produce the physical continuum limit. The number of
independent solutions is reduced leaving a continuum theory with phase
boundaries and a generally a non-trivial solution space. This process is very
complex and is carefully discussed in \cite{ggg;1996,ggzg;2007}. The important thing to
remember from this is that there are multiple solutions/phases to an
interacting QFT. This is ultimately the consequence of the non-linear nature of
the original interaction. In the early attempts to understand interacting QFT,
the only tool available was coupling constant perturbation theory built around
the free solution. The richness of the solution space of the theory could
not be observed and was not generally anticipated. Even now, the extent of
the possible solution space of quantum field theory is not fully appreciated.

It is possible to make the same comments about the solution space of QFT
starting with the Feynman path integral formulation. However, in most works
the path integral is defined as integration over fields valued on the real
axis. This does not yield the the full range of solutions of canonical field
theory that we discussed above. In fact, defined this way, the path integral
only produces solutions that are regular as the coupling vanishes. Such solutions
consequently approach those produced by coupling constant perturbation theory. The
conventional Path Integral formulation is correct as far as it goes, but it
excludes explicit access to much of the interesting content of canonical QFT,
such as symmetry breaking. However, many of these deficiencies are
corrected by extending the definition of the path integral into the complex
plane by not restricting the path integrations to the real axis but instead
allowing any range of integration through the complex plane for which the the
endpoints gives zero contribution. This procedure reproduces the Schwinger
action principle conclusions discussed above. Of course, the resulting complex
contributions must be combined so that only real (Euclidean) $\mathcal{Z}$
results. These comments will be expanded by studying the simplified lower
dimensional version of the above equation in the next paragraphs.

Zero space-time dimension means that only one point exists and thus the
lattice equation becomes:

  \begin{equation*}
    g\, \frac{d^3 \mathcal{Z}}{d J^3} + m^2\, \frac{d \mathcal{Z}}{d J} = J\, \mathcal{Z} \; .
  \end{equation*}
  While loosing any space-time structure and thus the possibility of
  understanding all the interesting structure that occurs in the continuum
  limit, the above still maintains the non-linear nature of quantum field
  theory and the associated multiple solutions.  Calculating solutions is now
  straightforward.  While the finite dimensional case potentially has an
  infinite number of solutions before accounting for the collapse of the
  solution set, the current equation, representing ``zero dimensional QFT''
  only has three independent solutions. The solutions can be found easily by
  using series methods. Alternatively, solutions can be found by examining the
  integral representation:
    \begin{equation*}
      \mathcal{Z} = \int e^{-g\, \frac{\phi^4}{4} -
        \frac{m^2}{2}\,\phi^{2} + J\phi}\, \mathcal{D}\phi \; .
    \end{equation*}

    This clearly is the Feynman path integral of the action in zero
    dimensions, which (within a normalization constant) corresponds to the
    vacuum-to-vacuum matrix element of a zero dimensional quartic scalar field
    theory. The integrand is negative of the Euclidean action of this
    theory. In order to obtain the same information as is available through
    solution of the differential equation, this integral must be evaluated
    over all possible independent paths in the complex plane where the
    contributions at the end points of the paths vanish. Straightforward
    evaluation shows that there are three allowed independent paths in the
    complex plane corresponding to the three solutions to the differential
    equation. It is always possible to weight each of the three solutions with
    a complex number in order to produce three independent real
    solutions. This in exact correspondence to the comments above for the
    theory with space-time structure. The integrand has 3 stationary points
    which are $(J = 0)$ located at
    \begin{equation*} \phi = 0, \; \phi =
    \pm\Biggl\{\hksqrt{\frac{-m^2}{g}}\Biggr\} \; .  \end{equation*}
    It is easy to expand around these saddle points to discover
    asymptotic expansions for each of the three solutions.

    The single field expectation value $\phi (J)$ is defined as
 \begin{equation*}
      \phi (J) = \frac{d \mathcal{Z}}{d J} \; .
    \end{equation*}
From the integral representation it follows that:

 \begin{equation*}
      \phi(0) = \int \phi\, e^{- g\, \frac{\phi^4}{4} -
        \frac{m^2}{2}\phi^{2}}\, \mathcal{D}\phi.
    \end{equation*}
    This integral has three values corresponding to the three independent
    integration paths. For the path along the real axis (corresponding to
    expanding around the stationary point at $\phi =0$) we find
    $\phi(0)=0$. This associated asymptotic solution reduces to the usual
    perturbation expansion around $g=0$ for small $g$, and as a consequence
    the vacuum expectation of a single field (or any odd power of the field)
    vanishes for a vanishing source. In general, the symmetry under reflection
    of $\phi$ of the integrand of the above integral representation might
    suggest we look only at integration paths that respect this symmetry, but
    if we did so we would leave out two of the solutions to the differential
    equation. For the two other paths, which include imaginary points,
    $\phi(0)$ is not zero and the global reflection symmetry of the integrand
    is ``spontaneously broken''. In theories with space-time structure, the
    situation is more interesting and complex because we have breaking
    associated with a conserved local current. However, the observation that a
    theory has multiple solutions as a consequence of its original
    non-linearity remains an essential fact, and it follows that most of these
    solutions will have lower symmetry (symmetry breaking) than the manifest
    symmetry of the original action.  Direct calculation shows that these
    solutions diverge as $g$ becomes small. They must diverge, else they could
    be described by coupling constant perturbation theory and produce the same
    result as that of the previously discussed path corresponding to the
    saddle point at the origin.

    Despite the fact that we have examined a simple system with one degree of
    freedom, our observations carry over (with appropriate modifications) to
    higher dimensions. At most, one of these solutions has a finite value at
    zero coupling and that solution is approximated by an asymptotic expansion
    around $g=0$ corresponding to coupling constant perturbation
    theory. Should the perturbative (Taylor) expansion in $g$ about $g=0$ not
    be asymptotic but convergent, then it is the only solution, which, in
    general, cannot be the case. This comment corresponds to the content of
    Dyson's famous paper \cite{dyson;1952}. The other solutions are generally
    of the spontaneously broken type, meaning that they have non-vanishing
    expectation values associated with some Green's functions that vanish in
    the perturbative calculation. These Green's functions are singular for
    vanishing coupling. This is easily confirmed in the zero dimensional
    quartic example given above. Unlike the relatively trivial zero
    dimensional cases, the symmetry of the action that is ``broken'' by a
    choice of boundary conditions (i.e. vacuum state) is usually
    continuous. This is reflective of the more complex and rich set of
    solutions that occurs in higher space time dimensions. An important
    feature of the complexity of the higher dimensional solutions follows from
    the fact that while in the zero dimensional case solutions are totally
    unconnected phases with no way to move from one phase to another, in
    higher dimensions it is possible to have phase transitions through changes
    of parameters so that different phases can be associated with different
    values of a parameter set labeling the solutions. This is discussed in Ref
    \cite{ggzg;2007}. Finally it should be noted that there appear to be
    interesting cases in higher dimensions where even the perturbative
    solution does not exist. This appears to be the case in 4 space-time
    dimensions where it is believed that there are no non-trivial solutions
    that are regular as the coupling goes to zero and this even while the
    perturbation theory appears to exit, in the end it apparently just defines
    a non-interacting theory. Note that this is a much more complicated
    phenomena then the case of cubic interacting scalar field theory in any
    space-time dimension. Even in zero dimensions, this theory, while having
    solutions, does not have any which are regular as the coupling vanishes.

\section{Examples of the Goldstone theorem } \label{sec:3}

     I now explicitly examine some four (1 time, 3 space) dimensional quantum field theories
     with spontaneous symmetry breaking. I begin with the simplest possible
     free model to demonstrate the ideas associated with Goldstone's theorem but
     will end up examining some very interesting non-perturbative solutions of
     the type discussed above.

    Assume that at least one state of \emph{lowest} energy, the vacuum,
    $\ket{0}$, exists, and that the free relativistic scalar
    field is described by the operator action:

    \begin{align*}
      \int d^4x \bigg[\frac{(\pa_{\mu}\, \phi)\, (\pa^{\mu}\, \phi)}{2} - \frac{m^2\,
        \phi^2}{2} \bigg] .\\
       \end{align*}

    From this, the free one-field Green's function equation is:
    \begin{equation*}
      (-\pa^2 - m^2)\, \ipop{0}{\phi(x)}{0} = 0\; .
    \end{equation*}
    The requirement that the vacuum energy-momentum  $P^{\mu}$ generates space-time
    translations, combined with the requirement that the vacuum is an
    eigenstate of this energy-momentum operator, leads to the useful condition that
    \begin{align*}
      \ipop{0;n}{\phi(x)}{0;n} &= \ipop{0;n}{\phi(0)}{0;n} \equiv n. \\
      \intertext{The label $n$ is introduced in the states to
        explicitly categorize the vacua by their expectation values. Because
        the fields are translated by exponentials of $P^{\mu}$, it follows
        that} m^2\, n &= 0 \; .
    \end{align*}
    Thus, the one-field Green's function reduces
    to the requirement that
    \begin{align*}
      m^2 &= 0 \text{ or} \\
      n &= 0
    \end{align*}
    where $n$ can be chosen to be an arbitrary real number. Each choice for $n$
    yields a new independent vacuum state and as a consequence there is an
    infinite number of totally disjoint (inequivalent) sets of states
    producing identical disjoint (free) theories \cite{ggfu;1964}.  This is
    an elementary example of Goldstone's theorem
    \cite{Nambu:1961tp,Nambu;1961tp2,jg;1961,gsw;1962}. As
    pointed out previously, this theorem says that if a charge associated with
    a conserved current in a relativistic field theory does not destroy the
    vacuum, the theory must have massless excitations.

    The current in this example is
    \begin{align*}
      J^{\mu}(x) &= \pa^{\mu}\, \phi(x) \\
      \pa_{\mu}\, J^{\mu}(x) &= \pa^2 \phi(x) = -m^2\, \phi \\
      \wasytherefore\; \text{if } m^2 = 0 &\Rightarrow \pa_{\mu}\, J^{\mu}(x) = 0 \; \\
      \intertext{and the charge is}
      Q &\equiv \int d^3x\, \left(\pa^{0}\phi(\vec{x}, t)\right) \; .
    \end{align*}

    This charge (which is defined only when it appears in commutation
    relations as follows) does not destroy the vacuum.  From the
    canonical commutation relations:
    \begin{align*}
      i\, \comm{\pa^0\phi(x)}{\phi(y)}\bigg|_{x^0=y^0} &= \delta^{(3)}(x-y) \\ \\
      \Rightarrow\; i\, \comm{Q}{\phi(y)} &= 1 \\
      \wasytherefore\quad i\, \ipop{0;n}{\comm{Q}{\phi(y)}}{0;n} &= 1 \;. \\
      \intertext{It should be noted that this is consistent with}
      \frac{dQ}{dt} &= 0 \; ,
    \end{align*}
    despite the fact that the vacuum $|0;n\rangle$ is not an eigenstate of the charge $Q$.  This
    example, with no interaction, is particularly simple. It does not show the
    complex phase structure that always occurs with interacting fields and is
    even demonstrated in the preceding zero dimensional discussion.  However,
    it does illustrate the association of non-vanishing field expectations of
    translationally invariant symmetric theories with massless particles.

    Massless particles, particularly now that we know neutrinos have a
    small mass, seem to be limited to photons. After the
    Goldstone theorem was discovered, it was natural to ask how
    symmetry breaking could be used to explain the photon.  After
    Bjorken gave a talk (1963) at Harvard examining this possibility,
    my thesis advisor, Wally Gilbert, suggested that I look at
    Bjorken's proposed alternative four fermion model of quantum
    electrodynamics \cite{jb;1963}, which is a variant of the
    Nambu--Jona-Lasinio model with vector-vector interaction:
    \begin{equation*} g\, (\bar\psi\, \gamma^{\mu}\, \psi)\, (\bar\psi\,
      \gamma_{\mu}\, \psi) \; .  \end{equation*} The methodology of Bjorken's
    model seemed dangerous because the ``breaking'' requires that the
    conserved current that appears in the interaction has a non-vanishing
    vacuum expectation value. This picks a preferred direction in space-time,
    which seems to destroy Lorentz symmetry and therefore could result in
    solutions that violate relativistic invariance. However, breaking in this
    manner will respect relativistic invariance if matrix elements are
    consistent with the operator commutation relations involving the six
    generators of the Lorentz group. I showed this restriction can be met for
    the above interaction by constructing an iterative solution scheme with
    non-vanishing expectation of the current, consistent with all conservation
    constraints including the commutation relations of the Lorentz group. This
    demonstrated that the details of Bjorken's work were not quite correct,
    but, with some modification, his conclusion that this theory is physically
    equivalent to QED is correct \cite{ggff1;194,ggff2;1964}. The result is
    rather elegant in that the symmetry breaking parameter becomes
    ``calculationally inert'' in the sense that it is harmless, and is just
    passed through any calculation involving commutators. In the process of
    doing this, I set up what was probably the first well defined all-order
    relativistic expansion about a non-perturbative saddle point. This
    expansion, in composite vector propagator loops, is identical to the
    approach later named ``the large-$N$ expansion''.  The resulting
    electrodynamic equivalent theory is a surprise, because from coupling
    constant perturbation theory this interaction is well known to lead to a
    hopelessly divergent non-renormalizable expansion in the coupling
    constant.  The expansion I examined corresponds to an expansion about a
    different saddle point of the path integral than that for the coupling
    constant expansion, and leads to a solution in a symmetry breaking phase
    that shows the requisite small $g$ singularities.

    Said differently, the solutions of the Bjorken model, as well as
    the usual solutions of the Nambu--Jona-Lasinio model, correspond
    to alternative (i.e., non-coupling constant perturbation)
    solutions of a theory with a four fermion interaction.  The new
    solutions can be shown to be equivalent to those of normal
    perturbative quantum electrodynamics. However, in this case, the
    original theory does not have gauge invariance and the
    propagators, as directly calculated, correspond to normal
    electrodynamics in the Lorentz gauge. This theory explicitly has a
    massless particle behaving as a photon because of the Goldstone
    theorem. It is perhaps useful to point out that we have not shown
    that vector-vector four fermion interactions are identical to
    normal QED, but have shown that these two different operator
    theories share one solution of their otherwise different solution set.

    To put some of the above into more explicit form, the solutions of
    the Bjorken model under discussion are associated with symmetry
    breaking boundary conditions that follow from the requirement that
    \begin{align*}
      \ev{j^{\mu}} &= n^{\mu} \\
      j^{\mu} &= \bar\psi\, \gamma^{\mu}\, \psi.
      \intertext{Clearly, since}
      \comm{\mathfrak{J}^{\mu\, \nu}}{j^{\lambda}} &\propto \big[g^{\mu\,\lambda}\,
        j^{\nu} - g^{\nu\, \lambda}\, j^{\mu}\big]\;
    \end{align*}
    where $\mathfrak{J}^{\mu\, \nu}$ generates Lorentz transformations. This requires that
    \begin{equation*}
       \mathfrak{J}^{\mu\,\nu} \ket{0} \neq 0  \; .
    \end{equation*}
    The symmetry of the vacuum is therefore broken, and
    $j^{\mu}\ket{0}$ contains a zero mass (spin-one) particle --- a photon. It is
    important to note that we have constrained the
    solutions by requiring translational invariance
    imposed by constant expectation values independent
    of space-time.

    \section{Gauge particles and zero mass} \label{sec:4}

    Despite the fact that Schwinger had convincingly argued by that
    time that there was no dynamical reason for the photon to have
    zero mass \cite{sch1;1962}, from the arguments I gave about the
    Bjorken model, I thought that I could construct a symmetry
    breaking argument that would require massless photons in
    conventional QED. This argument was wrong and, fortunately,
    Coleman detected this in my (early 1964) thesis
    presentation. Needless to say, this did not make me happy but I
    should have known better. We did not socialize for some time, but
    he was right and the chapter was removed. Even after
    removing the offending chapter from my thesis, I was still sure I
    had missed something. Indeed, it turned out that this chapter had
    in it the seeds that led to the GHK work.

Another set of information generated at Harvard also turned out to be
directly relevant, although that was not appreciated until after we
understood the basics of symmetry breaking and gauge fields. Well
before my thesis was finished, I spoke with Gilbert about another
project. He was very interested in how a massive field theory of
spin-one made the transition to electromagnetism as its mass vanishes
and had done some interesting calculations illustrating the mechanism
involved. I told Dave Boulware about this and he, in turn, discussed
this work with Gilbert, and they wrote a nice paper containing the
original calculations and other related calculations \cite{bg;1962}.
They observed that, given a massless \emph{scalar} particle $(B)$ and
a massless \emph{vector} particle $(A^{\lambda})$ with the simple
``interaction'':
\begin{equation*} g\, A^{\lambda}\,
\left(\pa_{\lambda}B - g\,A_{\lambda}\right) \end{equation*} there
results a free spin-one field with mass {$g^2$}. That this simple
quadratic action describes a massive vector and nothing more can be
anticipated by counting the degrees of freedom and observing that $g$
carries the dimensions of mass (the Boulware--Gilbert model (BG) has
a conserved current and a trace of gauge invariance). We realized,
during the course of writing the GHK paper, that the BG model is
essentially identical to the leading approximation we had developed
for broken scalar electrodynamics.

 In summary, by 1962 it was understood that the striking four
 dimensional calculations of quantum electrodynamics have a zero-mass
 photon because of the structure imposed by perturbative iteration of
 the free zero mass electrodynamic theory. That a massless photon is
 not mandated in general, was demonstrated by two counter
 examples. The two dimensional Schwinger model \cite{sch2;1962} (QED
 in two dimensions) demonstrated explicitly that gauge theories need
 not have zero mass, a result which was confirmed by the BG model in
 four dimensions.  Of course, while interesting, both of these models
 are somewhat trivial and not of direct physical importance. It turns
 out that the reason finite mass is not a feature of just these models
 but is also realizable in physical, non-perturbative gauge theories
 can be understood in a general way within the the framework presented
 in GHK.  Looking back, all that was needed to describe the
 ``Brout, Englert, Guralnik, Hagen, Kibble, Higgs'' phenomenon'' was
 available at Harvard in 1962.

    I was very fortunate to have received an NSF postdoctoral
    fellowship, and like many of the other recipients at that time,
    wanted to use this as an opportunity to go to Europe. Not only was
    I very interested in gaining a new perspective on daily life and
    physics, but also was attracted by the fact that the high value of
    the dollar would make the stipend go very far compared to what it
    could do in the U.S. This fellowship could be used at any
    institution that would welcome its holder. My first choice was
    CERN which seemed like a very interesting and exciting
    place. Fortunately, my request to visit with this fellowship was
    turned down by Van Hove, in what seemed a very rude manner. After
    some serious thought and discussions, I decided that the best
    place for me to go was Imperial College (IC). I was aware of the
    beautiful work done on renormalization by Paul Matthews and Abdus
    Salam and particularly on symmetry breaking by
    Salam (with Goldstone and Weinberg)\cite{gsw;1962}. Furthermore,
    this choice would involve no major language difficulties. Lowell
    Brown, a Schwinger student who was just returning from his NSF
    fellowship, partially spent at IC, thought this was a good choice
    and gave me valuable wisdom on the desirability of avoiding
    freezing of outdoor plumbing pipes and the necessity of keeping
    Alfa Romeos idling at all times during cold stretches. I had just
    ordered an Alfa for Italian delivery, as he had previously done. At
    the time, these were relatively cheap and fun sports cars, even
    though their lack of reliability was beyond anything most of us
    have ever experienced.

    Fortunately, Paul Matthews sent me a friendly letter of invitation
    to IC in response to my request for a two year visit. Initially, I
    did not fully appreciate how very appropriate it was to choose
    IC. This choice turned out to be a major life altering
    decision. Indeed, my understanding of the history of IC, which
    would have served to further validate my decision, remained
    incomplete until recently, when Tom Kibble, after reading a draft
    of this note, gave me a detailed summary of the intellectual
    development at IC before my arrival which, when combined with our
    work, led to the unified electroweak theory. It is very
    interesting and, because of this, I include most of his note to me
    at this point. Some of his comments are a bit premature for my
    developing story, but any confusion should become resolved as the
    reader continues.

    Tom wrote: ``There had already been a great deal of discussion at IC of the
    possibility of symmetry breaking in gauge theories.  Salam was convinced,
    from a very early stage, that the ultimate theory would prove to be a
    gauge theory.  As I'm sure you know, his student Ronald Shaw developed the
    same model as Yang and Mills independently at the same time (1954), though
    it was never published except as a Cambridge University PhD thesis.  And
    of course Walter Gilbert was his student too.  Initially, the emphasis was
    on a theory of strong interactions, but that gradually changed and already
    in 1958 Salam and Ward published their first attempt at a unified gauge
    theory of weak interactions.  There were several later versions, including
    the one you mention in 1964.  It ran up against two major obstacles, of
    course --- parity violation and the need to give the weak bosons a large
    mass --- both obviously demanding some kind of symmetry breaking.  Weinberg
    and Salam discussed this problem at great length when Steve was here on
    sabbatical, with each other and with me and others.  There was of course a
    lot of discussion of approximate symmetries, in the context of strong
    interactions.  But certainly for the weak interaction case, the nicest
    possible explanation was obviously spontaneous symmetry breaking.  So it
    seemed clear that the big stumbling block was the Goldstone theorem, which
    is why they were so interested in studying its proof, to see if they could
    find any loopholes (they didn't of course), leading ultimately to the 1962
    paper with Goldstone.

    So I would say that as a group we were very much primed to see this as a
    key problem.  That was certainly why I was so interested when you first
    started discussing your ideas on the subject.''

   When I started my NSF postdoctoral fellowship at Imperial College
   at the beginning of 1964, I was certain that something
   interesting would happen with gauge theories and symmetry breaking. At
   IC, which, in retrospect, was arguably the best high energy theory
   place in the world at that time, I met a fantastic bunch of
   physicists. The ones I interacted with most were Tom Kibble,
   Ray Streater, John Charap, Paul Matthews, and Abdus Salam.

Even though I did not know the detailed history recounted by Kibble
above, I expected no questions to the basic assumption of the
possibility of spontaneous breaking of symmetry. The
Goldstone-Salam-Weinberg paper \cite{gsw;1962} seemed to ensure
that. Indeed, I naively could not even conceive that anyone could
justifiably question that this could happen.  I was soon to learn
that, while Harvard was relatively safe ground protected by
Schwinger's large (but indifferent) umbrella, the understanding of
field theory in most of our community was much different and for the
most part probably far less sophisticated. The idea that there was
even such a thing as symmetry breaking in field theory was not
universally accepted, even at IC. Ray Streater (who was, in the language of those times, an
axiomatic or constructive field theorist) told me that his peers did
not believe that symmetry breaking was possible. Streater and his
community were certainly very sophisticated, but perhaps too much
so. A lot of arguing, and my construction of the free model of
symmetry breaking (examined earlier in this paper), convinced him that
the axioms that led to disbelief in symmetry breaking were wrong. Well
after these discussions, he published a very nice paper on this matter
\cite{rs;1965} which was largely responsible for convincing his more 
rigorous group of theorists that symmetry breaking was possible.

My discussions with Streater reinforced my belief in the power of the
Goldstone theorem and that (combined with my obsessive belief that the photon
was massless for reasons more basic than the smallness of the coupling
constant) led me to write in April of 1964 a paper that was published in
Physical Review Letters (PRL) \cite{ggfu;1964}.  This paper, the precursor to
GHK, contains the simple free scalar example and a related argument for
quantum electrodynamics. The electrodynamic construction has very useful
content in relativistic gauges. However, this paper has a subtle error for interacting
electromagnetism in the radiation gauge that led to our full understanding of
symmetry breaking with gauge fields.

    I must make it clear that neither Tom nor Dick were in any way
    responsible for the error because I was not wise enough to have
    discussed the paper with them until after I submitted it. I had,
    unintentionally, effectively compartmentalized my thesis work and
    discussions with Streater from my more or less ongoing
    conversations with Hagen and my new direct conversations with Tom.
    During frequent lunches, consisting of vile hard boiled eggs in
    crumb wraps, unspeakable other options, and dessert and almost
    everything else covered with a yellow custard sauce, we discussed
    physics in general and, in particular, the apparent failure of
    Goldstone's theorem in solid state physics. Although I was
    woefully ignorant of anything about this, it did not initially
    bother me much, because these models were non-relativistic, but
    Tom was sure it was important and succeeded in convincing me that
    this was the case.  He was correct, but trying to understand in
    detail how this actually worked, after we had solved the
    relativistic case, held up the publication of the GHK paper by
    several months.

    This compartmentalization fortunately ended within days after the paper
    \cite{ggfu;1964} was sent to PRL. Refining my handling of gauges, quickly
    led to a complete understanding of the amazing structure of symmetry
    broken scalar quantum electrodynamics. Because of delays caused by the
    many postal strikes in Britain at the time and the peculiarities of IC's
    mail, this paper did not get to PRL until June 1st. This was long after we
    knew it was wonderfully wrong, but I thought it would not be printed until
    after I received the proofs or, more likely, the unfavorable referee
    reports. I intended to modify it at that stage. I was traveling when the
    proofs arrived at Imperial College, and a very accommodating John Charap
    saw the returned paper in the mail, proofread it, and sent it back to
    PRL. I remain embarrassed that the paper was published, but yet there is
    much that is correct about it and the proper finishing of the analysis in
    it gives an elegant overview of symmetry breaking and the solution set of
    gauge theories. To my knowledge, only a handful of researchers, including
    Streater and a famous independent reviewer who thought the paper was of no
    value on general grounds, has ever referenced this work. Amusingly,
    Streater's elegant paper got a lot more attention.

    \section{General proof that the Goldstone theorem for gauge theories
      does not require physical massless particles} \label{sec:5}

    In this section, I will quantify much of the foregoing discussion. The
    realization that my paper \cite{ggfu;1964} had an error (which was also
    caught by Dave Boulware) was in a rather amazing way the final key to our
    understanding that the assumptions of the Goldstone theorem are not
    necessarily valid in theories not showing manifest Lorentz invariance. It
    follows, as an exact statement, that symmetry breaking in a gauge theory,
    does not require physical massless particles. This is because these
    theories have valid representations in gauges such as the radiation gauge
    that are not manifestly covariant. In the following, I repeat the
    arguments of my PRL paper \cite{ggfu;1964}, but now correctly analyzed, to
    show what happens to the massless constraint of the Goldstone theorem in
    gauge theories. Here, as in GHK, I avoid similar but more complex
    arguments of QCD and confine the discussion to QED. For QCD, see
    \cite{twbk;1967} and for a comprehensive overall review see
    \cite{ghk;1968}. In QED, there is an asymmetric conserved tensor current

    \begin{equation*}
      J^{\mu\, \nu} = F^{\mu\, \nu} - x^{\nu}\, J^{\mu} \end{equation*}
 which is easily shown to satisfy

     \begin{equation*} \pa_{\mu}\, J^{\mu\, \nu} = 0. \end{equation*}

      Proceeding in the ``usual'' manner, it can be argued that the four charges given by

      \begin{equation*}
      Q^{\nu} = \int d^3x \left[ F^{0\, \nu} - x^{\nu}\, J^0\right]
      \end{equation*}
      are time independent, and that

      \begin{equation*} \frac{dQ^{\nu}}{dt} = 0 \end{equation*} is an
      immediate consequence of current conservation. However, the existence of
      a conserved charge depends on the assumption that the surface integral
      of the spatial current vanishes over a closed surface as that surface
      tends to infinity. In quantum field theory, this possibility is tested
      by evaluating matrix elements of the current commuted with other
      operators. It is easily seen that, in a manifestly covariant theory,
      causality always ensures that the surface integrals over spatial
      currents vanish. Without manifest covariance, there is no such
      guarantee. While physical results evaluated in any gauge in QED must be
      fully consistent with special relativity, gauge dependent quantities
      have no such restriction. If a gauge that is not manifestly covariant is
      chosen for the vector potential, matrix elements involving $A^{\mu}$
      will reflect this lack of covariance.  The radiation gauge
      $\vec\nabla\cdot\vec{A} = 0$ is obviously not manifestly covariant, and
      yet it has many advantages. In particular, canonical quantization in the
      radiation gauge is straightforward and non-physical degrees of freedom
      are not required. Using standard radiation gauge commutation relations,
      the asymmetric current defined above satisfies:

    \begin{equation*}
      \ipop{0}{ [Q^{k} ,\, A^{l}(\vec{x}, t)] }{0} \neq 0\; .
    \end{equation*}

    If $Q^{k}$ were time independent, we could conclude that the right
    hand side is simply a constant and that, by the Goldstone theorem,
    $A^{k}$ excites a zero mass particle, namely the photon. In the
    case of no current (equivalent to $e=0$), this is true and there
    it is a trivial Goldstone theorem for free electromagnetism
    consistent with the photon being massless.  However, direct
    calculation using spectral representations shows that this
    expression is time dependent for $e \neq 0$! What went wrong?
    Exactly what was discussed above! To emphasize this very important
    point, we repeat that the radiation gauge is not explicitly
    Lorentz invariant, and we therefore cannot use causality to prove
    that the above commutator or any commutator involving gauge
    dependent quantities vanishes outside a finite region of
    spacetime.  This means that, even though $\pa_{0} J^{0\, 0} +
    \pa_{k} J^{0\, k} = 0$, we cannot neglect surface integrals of
    $J^{0\, k}$. In other words, charge leaks out of any volume!

    This leads us to consider the proof of Goldstone's theorem with
    consideration for currents other than the special one introduced
    above. What we have learned is applicable to any current,
    and it follows that Goldstone's theorem is true for a manifestly
    covariant theory, namely a theory where $\pa_{\mu} J^{\mu} = 0$
    and surface terms vanish sufficiently rapidly, so that

    \begin{equation*} \ipop{0}{\comm{\int d^3x (\pa_{\mu}
    J^{\mu})}{\text{(local operator)}}}{0} = \frac{d}{dt}
    \ipop{0}{\comm{\int d^3x\, J^0}{\text{(local operator)}}}{0} \;.
    \end{equation*}

 Under these circumstances,
\begin{equation*} Q = \int d^3x\, J^0 \end{equation*}
has a massless particle in its spectrum. All the original proofs
assumed manifest covariance, so there was never the possibility that
the commutators involving an infinitely distant space-like surface
could contribute. If the theory is not manifestly covariant, there is
no guarantee that the ``charge'' is effectively time independent in
all commutators.  In particular, Goldstone's theorem need not and does
not require physical massless states in any gauge theory.  This is
because these theories are made to be manifestly relativistic through
the introduction of extra (i.e., gauge) degrees of freedom. Indeed,
the Goldstone bosons are always non-physical.  Consistent with this
observation, when the Green's function
$\ipop{0}{\comm{Q^{k}}{A^{l}(\vec{x}, t)}}{0}$ is re-gauged to the
manifestly covariant Lorentz gauge, it becomes a non-zero constant,
and the conditions of the Goldstone theorem are met. The resulting
massless excitations are pure gauge.

Before moving on, it is important to make some additional observations.  In
the trivial case when $j^{\mu} = 0$ in the above example, the Goldstone
theorem is valid and the free electromagnetic field does indeed have zero mass
associated with a legitimate Goldstone theorem.  The free case is, however,
fundamentally different from the interacting case, for which there is no
Goldstone theorem (even though perturbation theory does yield a massless
photon). Thus, despite the absence of a Goldstone theorem for physical
particles, the photon maintains its masslessness, as argued by Schwinger,
because of the smallness of the renormalized coupling constant. The
perturbation solution is just one of a set of solutions possible for small
coupling as a consequence of the non-linearity of the equations of
motion. This is a manifestation of Dyson's argument \cite{dyson;1952}, which
shows that the perturbative solution is asymptotic when evaluated around
vanishing coupling constant. (It is interesting to note that the exact
solutions for zero dimensional models, such as the quartic one discussed
above, confirm the asymptotic behavior deduced by Dyson.) If the Taylor
expansion in $e$ existed, the solution would be unique. The other, totally
independent, solutions are associated with symmetry breaking, and must be
singular in the $e\rightarrow 0$ limit. Even though there is no Goldstone theorem
for the electromagnetic interaction, the perturbative solution has a
massless photon. It is necessary to verify that other
solutions are not massless and actually are associated with massive vector particles.  The GHK
paper provides this verification by looking at the simple example of broken
scalar electrodynamics.

\section{Explicit symmetry-breaking solution of scalar QED in leading order,
  showing formation of a massive vector meson} \label{sec:6}

We now construct the classical example of the failure of Goldstone's theorem
by considering the action \begin{align*} L &= -\frac{1}{2}\, F^{\mu\, \nu}\,
  (\pa_{\mu}A_{\nu} - \pa_{\nu}A_{\mu}) + \frac{1}{4}\, F^{\mu\,
    \nu}\,F_{\mu\, \nu} + \phi^{\mu}\pa_{\mu}\phi + \\ &\quad+ \frac{1}{2}\,
  \phi^{\mu}\phi_{\mu} + i\,e_0\, \phi^{\mu}\, \mathbf{q}\, \phi\,
  A_{\mu} \\ \mathbf{q} &= \sigma_2 \\ \phi &= (\phi_1, \phi_2) \\
  \phi^{\mu} &= (\phi^{\mu}_{1}, \phi^{\mu}_{2}) \; .  \end{align*} This is a
useful way of writing scalar electrodynamics in terms of real fields.  Observe
that we have not added any explicit self interactions of the scalar boson
fields or an explicit scalar bare mass.  This is also the case in the GHK
paper. This was done intentionally so as to put the emphasis on what happens
to the vector particle. To leading order in the symmetry breaking
approximation we study, additional scalar terms (such as a scalar mass term or
quartic scalar interaction) added to this action have no effect on the vector
field. In GHK, we wrote down the field equations that follow from the above
action. As we were well aware, the cleanest way to proceed is to add sources
to the action, then to write the Green's functions that follow from the field
equations, and then define an iterative approximation scheme. This is how I
handled such problems in my thesis and associated papers, and indeed in notes
building up to the GHK paper. (Note, that the expansion generated in this case is
not equivalent to the large-$N$ expansion.) However, since we chose the
leading order approximation to be a linearization of the operator equations,
it was, in this case, reasonably simple to construct an internally consistent
approximation without invoking the entire Green's function methodology. We
felt this was the clearest and most efficient approach for a space-limited
journal such as PRL.

   In the next paragraphs, we outline the elements of the GHK
   calculation performed in the radiation gauge. Our scalar
   electrodynamics model has two physical field degrees of freedom
   associated with the electromagnetic field and two physical field
   degrees of freedom associated with the spinless boson field along
   with their corresponding conjugate momentum fields. We solve this
   theory in the symmetry breaking phase by imposing the symmetry
   breaking condition

\begin{equation*}
  i\, e_0\, \mathbf{q}\, \langle 0| \phi | 0\rangle = \eta =
  \begin{pmatrix}
    \eta_1 \\
    \eta_2
  \end{pmatrix}\; .
\end{equation*}
The leading approximation is obtained by replacing $i\, e_0\, \phi^{\mu}\, \mathbf{q}\, \phi\, A_{\mu}$ in
the Lagrangian by $\phi^{\mu}\, \eta\, A_{\mu}$.
(The result is essentially the Boulware-Gilbert action \cite{bg;1962} with an extra scalar field)

This ``reduced Lagrangian'' results in the linearized field equations:

\begin{align*}
  F^{\mu\, \nu} &= \partial^{\mu} A^{\nu} - \partial^{\nu} A^{\mu} \; ;\\
  \partial_{\nu} F^{\mu\, \nu} &= \phi^{\mu}\, \eta\; ; \\
  \phi^{\mu} &= -\partial^{\mu}\phi - \eta\, A^{\mu} \; ; \\
  \partial_{\mu}\phi^{\mu} &= 0\; .
\end{align*}

These equations are soluble, since they are (rotated) free field
 equations. The diagonalized equations for the physical degrees of freedom
 are:

\begin{align*}
  (-\partial^2 + \eta_1^2)\, \phi_1 &= 0 \; ;\\
  -\partial^2 \phi_2 &= 0 \; ;\\
  (-\partial^2 + \eta_1^2)\, A_k^T &= 0 \; .
\end{align*}
For convenience, we have made the assumption that $\eta_1$ carries the full
value of the vacuum expectation of the scalar field (proportional to the
expectation value of $\phi_2$).  The superscript ${}^T$ denotes the transverse
part.  The two components of $A_k^T$ and the one component of $\phi_1$ form
the three physical components of a massive spin-one field while $\phi_2$ is a
spin-zero field. As previously mentioned, the Goldstone theorem is not valid,
so there is no resulting massless particle. If the Goldstone theorem were
valid, $\phi_1$ would be massless. It is very important to realize that it is
an artifact of the lowest order approximation for the above action that
$\phi_2$ is massless. The excitation spectrum of this field is not constrained
by any theorem. I emphasize that the Goldstone theorem naively (and
incorrectly) appeared to constrain $\phi_1$ to have a massless excitation but
neither it or any other condition imposes a direct constraint on the spectrum
of $\phi_2$. $\phi_2$ acquires mass if higher order corrections are calculated
for the simple interaction above or in leading order if scalar
self-interactions of $\phi$ are added to our simple action. Of course, the conclusions
about the invalidity of the Goldstone theorem are unchanged. An interesting
example of this results from the addition of an explicit quadratic scalar bare mass term and
a scalar quartic interaction term. In this case, in lowest order, the electromagnetic field
has no effect on $\phi_2$ and the equations satisfied by $\phi_2$ and hence its
leading order mass are identical to those determined by Goldstone in his
original (purely scalar field) symmetry breaking paper \cite{jg;1961}. The
boson excited by $\phi_2$ has come to be called the Higgs boson. The equations
and hence the results that lead to the massive spin one particle in our
original example are unchanged.

At this stage, it might be thought that we have written down an
interesting, but possibly totally uncontrolled, approximation.  There
is no \emph{a priori} reason to believe that this is even a meaningful
approximation. The main result, that the massless spin-one field and
the scalar field unite to form a spin-one massive excitation, could
be negated by the next iteration of this approximation. However, this
approximation meets an absolutely essential criterion that makes this
unlikely. While the symmetry breaking removes full gauge invariance,
current-conservation, which is the fundamental condition, is still
respected. This is clear from the above linearized equations of
motion. Further, we can directly demonstrate that the mechanism,
described earlier in this note for the failure of the Goldstone
theorem, applies in this approximation. As shown by the above
equations, the linearized conserved operator current is $J^{\mu} =
\phi^{\mu}\,\eta$. Using this, direct calculation leads to the result
\begin{equation*} \begin{split} \ipop{0}{\int d^3x\comm{J^0(\vec{x},
t_1)}{\phi_2(\vec{y}, t_2)}}{0} &= \eta_1\, e^{-i\,\eta_1\,(x^0 -
t^0)} \\ &\neq \eta_1 \end{split} \end{equation*} so that the charge
varies in time and is therefore not conserved, in accord with our
claim. This is a direct demonstration of the failure of the Goldstone
theorem.  Observe that the zero-time value of this commutator yields
$\eta_1$, which is proportional to the vacuum expectation value of
$\phi_2$, consistent with the equal time commutation relations. The
internal consistency and the consistency with exact results gives this
approximation credence as a leading order of an actual solution. As
mentioned previously, it is, in fact, not hard to make this the leading
order of a well defined approximation scheme.

Historically, this entire set of results existed in the spring of 1964. The
only part of the published GHK argument that was missing was the detailed
calculation of the time variation of the charge as shown in the above
equation. It seemed clear that our arguments, as presented above, would
explain why the Goldstone theorem could not be expected to be valid in
condensed-matter, but we lacked confirmation through detailed
calculation. Because of my early discussions with Kibble had often brought up
this question, I felt that we had to do a condensed-matter calculation
before publication.

   I spent the summer of 1964 traveling through Europe (on \$5.00 a
   day) and stopped to visit my advisor Walter Gilbert in Italy, where
   he was giving biology lectures at a meeting at Lake Como.  I
   explained to him how it all worked and how close we had been a year
   before. He had published a clever paper in PRL \cite{wg;1964}
   earlier that year that still missed the point, but nevertheless
   came close to describing aspects of a mechanism to avoid massless
   particles. The work that I showed Gilbert was complete in every
   major way and identical to that in the GHK paper, but for the
   beautiful consistency check of the charge oscillation in the
   leading order approximation that Hagen did in late summer.

   We had no idea or suspicion that anyone else was working along
   related lines and felt absolutely no sense of urgency to publish
   our results even though I had been freely discussing them with
   anyone who would listen. From the reactions I got, I thought it
   unlikely that anyone else would ever be truly interested in our
   approach, let alone believe the results.

   After I returned in late August, I made several visits to Oxford to
   talk to my old Harvard friend Bob Lange, to see if we could figure
   out how this worked in condensed matter.  We did not succeed at
   that time but Lange figured it out later and published in PRL
   \cite{rvl;1965} in January 1965.

   After giving up on the condensed matter problem, I conceded that it
   was time to publish.  Hagen had arrived at IC for a visit thus
   making collaboration straightforward. We cleaned up the arguments
   and he did the charge oscillation calculation in the approximate
   model. This calculation confirmed the general argument and
   convinced us that our conclusions were ironclad. After he finished
   this and confirmed all the other results, I was confident that,
   this time, there were no blunders. I had learned over the years,
   that Dick was impressively accurate and rarely if ever missed an
   error.

   Dick and I wrote up the paper and we gave it to Kibble for final
   scrutiny. As we were writing, he provided the essential insight and
   wisdom that insured that we had a coherent and logical argument. I
   certainly would have quit after my second stupid attempt to prove
   the photon massless but for Tom's depth of understanding of what
   was going on. Tom approved the paper.

  \section{Our initial reactions to the EB and H papers} \label{sec:7}

  Shortly thereafter, as we were literally placing the manuscript in the
  envelope to be sent to PRL, Kibble came into the office bearing two papers
  by Higgs and the one by Englert and Brout. These had just arrived in the
  then very slow and unreliable (because of strikes and the peculiarities of
  Imperial College) mail. We were very surprised and even amazed. We had no
  idea that there was any competing interest in the problem, particularly
  outside of the United States. Hagen and I quickly glanced at these papers
  and thought that, while they aimed at the same point, they did not form a
  serious challenge to our work.

  Higg's Physics Letters paper \cite{phpl;1964} indicates that there might be
  an escape from the masslessness condition imposed by the Goldstone theorem,
  if calculations are done in the radiation gauge. However, we felt that an
  explicit quantum field theoretic example is needed to show that this had
  content. This follows from the result outlined above where I show the
  ``failure'' of a Goldstone theorem in pure unbroken radiation gauge
  electromagnetism. This happens through essentially the same mechanism as
  outlined by Higgs in his paper. Nevertheless, perturbative QED is
  characterized by a physical zero mass photon. In the end, to obtain the
  results necessary for the unified electroweak theory, it is necessary to
  show that there is no physical zero mass excitation.  Higgs attempts to fill
  in this deficiency in his PRL paper \cite{ph;1964}, but does not revisit the
  radiation gauge and does not completely calculate the spectrum in this paper
  as discussed in more detail below.

  We felt that Englert and Brout's work \cite{eb;1964} was clearly related to
  GHK, but presented a less than completely defined approximation. This paper
  (as well as the Higgs PRL paper discussed below) did not appear to fully
  recognize how very important it is to keep track of degrees of freedom. As a
  result, these authors did not provide, nor correctly comment on, the entire
  mass spectrum of their models. In particular, EB ``assumed'' that the
  Goldstone theorem is correct, which is true in their case, since they
  calculate in a covariant gauge. However, they demoted the corresponding
  massless excitation to a ``pole at $q=0$ which is not isolated.'' The
  correct leading order approximation reveals a distinct zero mass pole. This
  pole is the one required by the Goldstone theorem. From the detailed
  calculation is easily seen that is purely gauge in nature. That is to say,
  this pole does not contribute to any physically measurable quantity. EB does
  not do the calculation or make this observation. While EB does start with a
  two component scalar field, no comment is made on the spectrum associated
  with the component with non-vanishing expectation value. This component
  provides the ``Higgs Boson.'' To sum up, we felt that EB had found the
  dimensional parameter created by symmetry breaking and used it to make a
  massive boson. This, in itself, is only part of the problem. They did not
  provide a convincing argument to justify the correctness or consistency of
  their approximate and partial lowest order symmetry breaking solution.

  Higgs' more complete PRL paper \cite{ph;1964} examines an approximation to
  broken scalar electrodynamics but does not pick a gauge.  As I have pointed
  out, a consistent calculation in a covariant gauge must have a massless
  Goldstone boson and to establish if this excitation is a gauge particle or
  an actual particle takes analysis. In the radiation gauge, there is no
  Goldstone theorem, but the explicit absence of a physical massless modes
  must be confirmed by direct calculation. Higgs missed this entirely. He
  proceeded by writing down an approximation to the exact arbitrary gauge
  equations of motion. He then observes that these approximate equations of
  motion describe a massive vector particle and a scalar particle with mass
  determined in this approximation by the form of the initial scalar
  interaction. No statement is ever made about gauge. While the physical
  content of this coincides with the end result of GHK, much is taken for
  granted in reaching this conclusion, and it is not justified or developed in
  detailed solutions using covariant gauges. The whole analysis is classical.
  As it has been stated, the Goldstone theorem requires quantum mechanics and
  hence quantum mechanics is essential to describe accurately and fully the
  phenomena that occur. While Higgs deferred to Englert and Brout for quantum
  mechanics, as mentioned, we felt that this work was less than complete. The
  quantum structure and symmetry structure are determined by the exact
  equations. If the leading approximation is not consistent with these
  conditions then some additional ``higher order'' approximation must correct
  these inconsistencies. This means that any result from the so called leading
  approximation is likely to be negated by corrections. That is to say the
  approximation, if not completely consistent with the general requirements of
  the exact equations, can not be expected to predict actual physical results.
  Higgs did not subject his approximation to this test in that he failed to
  observe that his approximation is gauge sensitive. In this approximation a
  careful calculation in radiation gauge shows no massless particle while the
  calculation in a covariant gauge does have a massless particle to satisfy
  the Goldstone theorem but this particle is purely gauge in nature.

  Neither EB or Higgs fully analyze the consistency of their approximations
  nor recognize that massless Goldstone Goldstone particles survive in
  covariant gauges but are unphysical gauge particles. This argument is a
  center piece of the GHK calculation. I quote my collaborator C.R. Hagen,
  ``In a sense EB and H solved half of the problem --- namely massifying the
  gauge particle.  GHK solved an entire problem --- massifying and also showing
  how the deadening hand of the Goldstone theorem is avoided.''

  In summary, we felt that while these papers aimed in the correct direction,
  they did not form the basis for serious calculation. Because of the many
  discussions we had with those outside our collaboration, we knew that our
  work was going to be very controversial (i.e., generally regarded as just
  plain wrong), and the EB and H approaches even more so. Once there is a
  dimensional parameter from the symmetry breaking, it is easy to put in a
  mass almost anywhere you want it. It is quite another thing to show that you
  have a self consistent theory that could be the basis for an extension
  beyond the initial approximation. Although these were observations made in
  haste and with the arrogance and exuberance of youth, I still feel,
  particularly with the understanding and concerns of that time, that they
  were essentially correct.

   At the same time, Kibble brought our attention to a paper by P.W. Anderson
   \cite{pwa;1963}. This paper points out that the theory of plasma oscillations is
   related to Schwinger's analysis of the possibility of having relativistic
   gauge invariant theories without massless vector particles. It suggests the
   possibility that the Goldstone theorem could be negated through this
   mechanism and goes on to discuss ``degenerate vacuum types of theories'' as
   a way to give gauge fields mass and the necessity of demonstrating that the
   ``necessary conservation laws can be maintained.''  In general these
   comments are correct. However, as they stand, they are entirely without the
   analysis and verification needed to give them any credibility. These
   statements certainly did not show the calculational path to realize our
   theory and hence the unified electroweak theory.  It certainly did not even
   suggest the existence of the boson now being searched for at Fermi lab and
   LHC. The actual verification that the same mechanism actually worked in
   non-relativistic condensed-matter theories as in relativistic QFT had to
   wait for the work of Lange \cite{rvl;1965}, which was based on GHK. We did not
   change our paper to reference the Anderson work.

   In any event, after seeing the competing EBH analyses, we unhesitatingly
   thought that we should do the completely honest thing and reference them,
   as they were clearly relevant with examples, even if not convincing to us.
   Our paper was finished and typed in final form when we saw these other
   works and made this decision. We only altered the manuscript by adding in
   several places references to these just-revealed papers. Not a single
   thought or calculation was removed or added, nor was any change made but to
   the referencing in our manuscript as the result of Kibble's having pointed
   out the existence of these new papers. In retrospect, I wish we had added
   the true statement --- ``after this paper was completed, related work by EB
   and H was brought to our attention.''

We were naive enough to feel that these other articles offered no threat to our
insights or to the crediting of our contribution. Nearly 45 years later, it is
clear that we were very wrong.  An unbiased reading of all the papers should
make it clear that GHK is the result of an entirely independent train of
thought. Over the decades, the awareness of the need to address what in 1964
were many worrisome points has vanished because of the acceptance of the end
results and the general increase in our theoretical understanding. This, in
turn, has affected the appreciation of the extremely significant differences in
``correctness'' and ``completeness'' of our work relative to the others. While we were
too innocent in our slowness to publish and in the way the referencing was
included, we never thought that this would in any way affect our
claim, since so many knew of the evolution of our work and my open discussion
of it for nearly six months before we submitted it for publication.  We were
sure that a clear claim had been staked for us, even without the
first-publication place holder. Over the short term this was correct, but,
after decades, colleagues died and many seemed to forget.  A major loss in this
regard and many many other ways was Paul Matthews, who was aware of the
detailed evolution of our work. Initially, there seemed to be no problem
getting recognition for what we did on a more than equal basis to the EB and H
papers. This seemed to change around 1999, when our work began to be omitted
from the references contained in important talks and papers, even by authors
who had previously referenced us.

   \section{Reactions and thoughts after the release of the GHK
     paper} \label{sec:8}

   What followed after the paper was sent out is quite
   interesting. While I spoke about this work informally to many
   people and in many places before the GHK paper was released, I also
   gave several seminars after its release. My presentations were
   greeted with fairly uniform disbelief. I was told in no uncertain
   terms that I did not understand electromagnetism or quantum field
   theory. In a community conditioned by coupling constant
   perturbation theory, it seemed that our work was nonsense. It is
   probably interesting to note that all of my talks were in Britain
   or Europe. After I moved to Rochester in the fall of 1965, I was
   never again asked to speak about GHK, not even at Brown, until the
   colloquium that I gave in 2001 at Washington University, on which
   this paper is based. Similarly, Hagen has never given a talk on our
   paper other than at Rochester, while I believe that Kibble spoke at
   several institutions.

   Two particularly interesting talks were ones I gave at Edinburgh and at a
   conference outside of Munich.  The work on symmetry breaking done by Higgs
   caught the eye of N. Kemmer, who was the professor of theoretical physics
   at Edinburgh. He wondered what his colleague was up to, and called Paul
   Matthews (who was Kemmer's student) at Imperial College. Paul, who was
   always very kind to me, told Kemmer that he should invite me to speak at
   Edinburgh, and see if that helped him make head or tails about what was
   going on. My wife and I visited Edinburgh on November 23, 1964. I gave a
   seminar and had a delightful time talking with colleagues and particularly
   Peter Higgs.  In the evening, we had a pleasant dinner with him and his
   wife. I found Peter to be a very warm and friendly person. I recall
   thinking that his understanding of the topic of symmetry breaking was less
   extensive than ours, and I offered him my version of how everything fit
   together. He published much of that discussion (with acknowledgment) in his
   1966 Phys.Rev. article \cite{phr;1965}.

   In the summer of 1965, I gave a talk at a small conference outside
   of Munich, that was sponsored by Heisenberg \cite{hc;1965}. He and
   the many other senior physicists at the conference thought these
   ideas were junk, and let me know with much enthusiasm that they
   felt that way. This evaluation, was made very clear to me by
   Heisenberg, who arguably had discovered spontaneous symmetry
   breaking in the first place. This contributed considerably to my
   fear that I could not survive in physics. Ken Wilson also spoke at
   this conference on his ideas of doing calculations on space time
   lattices. He also got beaten up rather badly. Hagen spoke (twice)
   at the same conference but on different topics.

   One redeeming aspect of this conference was that I got a demonstration
   ride in Julian Schwinger's factory-fresh Iso Rivolto (a beautiful quick
   machine that was powered by a Corvette engine). Julian remembered from my
   Harvard days that I loved cars and would
   be very interested in the wonderful machine on which he had spent a
   noticeable part of his Nobel prize money. The ride was made all the more
   interesting by Mrs. Teller attaching herself to us as we walked to the
   car. She sat in the front seat, thereby placing
   me in the ``imagination seats'' in the back. While I struggled to breathe, she
   told poor Julian that ``in the US such expensive cars have automatic
   transmissions''. This was uttered while he was in the midst of a stunning display of
   clutch work. Schwinger was kind enough not to say a word about my talk, even
   after the ride was over and Mrs. Teller had left.

   As formidable as I found Mrs. Teller, her famous husband impressed and
   scared me even more. When I was sitting alone at a table in the hotel where
   the conference was held, Teller sat down and asked me to explain $SU(6)$
   classification schemes, which were popular at that time. Fortunately,
   although it was not my thing, I had a decent knowledge of the
   literature. Teller grilled me without mercy for what seemed like
   hours. When he was finally satisfied, he left.  Despite my efforts to make
   myself invisible, he caught me again the next day and showed me a large
   number of calculations that he had performed. They were extensive in their
   coverage of the subject and, where I knew the results, absolutely correct.

   My experiences in general at this conference, the first one at which I
   spoke, left me feeling depressed and more than a bit beat up and worried
   about my survival as a physicist. Fortunately for me, Hagen helped me
   get a postdoctoral job at Rochester, where Robert Marshak was the dominant
   force, as well as the head of the high energy theory group.  Marshak, who
   was a commanding and wonderful presence, had a conversation
   with me after I had been at Rochester for about a year. Much of what I was
   doing involved symmetry breaking and was done with Dick. He told me I had
   to work on something else if I wanted to stay in physics.  The job market
   was very tight (not a new thing!).  I obeyed. I am still sure he was
   correct.

   Years later, at the 2nd Shelter Island conference, he publicly apologized to
   me for stopping my work on symmetry breaking and ``probably stopping him
   from getting a Nobel prize.'' There were many important people present, and
   I remain impressed by his decency and courage as well as his excessive
   faith in my abilities.

   \emph{``What about the unified theory? How did we miss it?''} Timidity,
   slowness and bad luck. After the GHK paper was published, John Charap and
   I, while sitting in his Ford Anglia in a rainstorm, had a discussion about
   the possibility of describing weak interactions unified with E\&M through
   this mechanism. We thought it was possible, but the idea drifted away,
   largely through lack of action on my part.  I dismissed the possibility of
   working on it because of other interests and because I was mostly receiving
   a less than warm response on GHK. Once again, I foolishly did not report my
   conversation with Charap to Dick or Tom. I did not focus seriously on
   this idea again until I went to Rochester. I started thinking about this
   discussion while at Rochester because of Marshak's intense interest
   in the weak interactions. This was clearly an interesting possibility. But
   because Hagen was away at the time, I kept my thoughts to myself and let
   them slip entirely after my ``survival'' discussion with Marshak.

   Another bit of bad luck came about earlier at Imperial College in my
   interactions with John Ward. Around the same time that we were working on
   symmetry breaking with gauge fields, Salam and Ward were working on a
   precursor to the Weinberg-Salam model. They were rather secretive about
   this, but one day a case of champagne appeared at the Imperial College
   physics department. I was told this was in anticipation of the prize they
   were going to get for their current work. Shortly after this, Ward and I
   went to a pub together for lunch. I started to tell him about our work on
   symmetry breaking but did not get far before he stopped me. He proceeded to
   give me a lecture on how I should not be so free with my ideas because they
   would be stolen and often published before I had a chance to finish working
   on them. Needless to say, I did not ask him about his work with Salam! If
   he had only listened, the two of us had enough information to have had a
   good chance to solve the unification problem on the spot. Of course, I
   could have read their papers after they came out, but I did not do
   this. Ironically, later Tom explained the details of our work to Salam who
   used this information to complete his unification model. This was told to
   me by Tom and is verified in the body of Salam's Nobel acceptance
   speech. To sum it up, I was simply not paying attention to all the signals
   coming to me. In hindsight, they were clear.

   While I was at Rochester, I got several calls from my Harvard classmate and fellow Gilbert student
   Marty Halpern, who was and still is at Berkeley. He asked me many questions about
   our paper and told me that he would be passing on the contents of our conversation  ``to
   Steve.'' I would like to think that this helped Weinberg put it all together
   for his brilliant paper \cite{sw;1967}, but I have no idea if any of the conversations were
   actually passed on. I had already stopped thinking about symmetry breaking
   because of Marshak's warning.

   In retrospect, my work with symmetry breaking was really fun and
   exciting. As mentioned at the beginning, I made new friendships,
   particularly with Tom and his, very sadly deceased, wonderful wife
   Anne. Dick and Tom (and others, particularly at IC) taught me much
   about how to think and how to be a practicing physicist. I made
   many errors of judgment and certainly errors in physics. Facing up
   to the possibility of errors and particularly career-damaging ones
   was hard for me, and surely made me be more conservative than I
   should have been. At this stage, the ideas seem very simple and
   natural. At the time they were not.

\section*{Acknowledgments}
I wish to thank Richard Hagen and Tom Kibble for the collaboration that made
the GHK paper possible, as well as for discussions over the years that have
taught me so much. I very much appreciate their careful reading and correction
of this manuscript. I am very grateful to Tom Ferbel for encouraging me to
finally put this historical story in print, and for careful and important
corrections for clarity and style as well as physics content. Thanks go to my
students Daniel Ferrante and Cengiz Pehlevan and to my son Zachary Guralnik
for many observations that have made this a more accurate and easier to read
article.  I have also benefited from informative discussions with Roman
Jackiw. This work is supported in part by funds provided by the US Department
of Energy (\textsf{DoE}) under \textsf{DE-FG02-91ER40688-TaskD}.

\end{document}